# StegoHound: A Novel Multi-Approaches Method for Efficient and Effective Identification and Extraction of Digital Evidence Masked by Steganographic Techniques in WAV and MP3 Files

Mohamed C. Ghanem [a & b], Maider D. Uribarri [a], Ramzi Djemai [a], Dipo Dunsin [a] and Istteffanny I. Araujo [a]

[a] *Cyber Security Research Centre, London Metropolitan University, 188-220 Holloway Road, London, N7 8DB, UK*

[b] *Department of Computer Science, University of Liverpool, Liverpool L69 3BX, UK*

## Abstract

Anti-forensics techniques particularly steganography and cryptography have become increasingly pressing issues that affect the current digital forensics practice. This paper advances the automation of hidden evidence extraction in the context of audio files by proposing a novel multi-approaches method which enables the correlation between unprocessed artefacts, indexed and live forensics analysis and traditional Steganographic and Cryptographic detection techniques. In this work, we opted for experimental research methodology in the form of a quantitative analysis of the efficiency of the proposed automation detecting and extracting hidden artefacts in WAV and MP3 audio files by comparing it to standard industry systems. This work advances the current automation in extracting evidence hidden by Cryptographic and Steganographic techniques during forensics investigations, the proposed multi-approaches demonstrated a clear enhancement in terms of coverage and accuracy notably on large audio files (MP3 and WAV) for which the manual forensics analysis is complex, time-consuming and requires significant expertise. Nonetheless, the proposed multi-approach automation may occasionally produce false positives (detecting steganography where none exists) or false negatives (failing to detect steganography that is present) but overall achieve a good balance between efficiently and effectively detecting hidden evidence and minimising the false negative which validates its reliability.

Corresponding Author: Dr Mohamed C. Ghanem (ghanemm@staff.londonmet.ac.uk)

## 1. Introduction

In the current era, life is considered incomplete without the Internet, as everyone has some data to transmit and receive. Data secrecy and privacy play a vital role in transferring sensitive data over the internet. In a world in dire need of technology, thousands of innovations take place everywhere and all the time, not just in protection, but in criminality. With the rapid advancement of technology in every field, people demanded high-performance communication platforms. Many individuals claim that technology has transformed the world into a global interconnected community. Various devices and platforms are required to communicate over the Internet and communication played a crucial role in the modern era, where people wanted to reach one another rapidly independent of location [36]. Communication comes from the start of human life; one could even say it is one of the basic needs of people. People have introduced a variety of services based on basic mechanisms to communicate over the Internet. Individuals over 30 years old working and playing in this field can recognise the fast development and changes applied. With rapid adoption, where a sender and receiver are connected through a medium to exchange messages, the communication became a target for attackers looking to disrupt communication patterns for various purposes. Different attacks and security breaches have been carried out by malicious users, but the communication medium was less affected as compared to the direct information attack where information is either directly taken or manipulated by the intruders. The information transmitted over the internet from sender to receiver is taken by the attacker using various attacks like phishing, spoofing, sniffing and man-in-the-middle attacks. Consequently, numerous security-based mechanisms were introduced and implemented to make secure communication possible. Firstly, the information hidden by using text shifting and substitution methods failed to provide a prominent level of security. Such techniques were easy to break because of the static key set by the sender. With the enhanced popularity and demand for security, an old, but improved technique persisted in the communication world which we know as encryption and has various new forms and algorithms today [1]. Encryption is an important method for people and organisations to protect sensitive data from hackers.

Encryption is a technique where the text is transformed or converted into a code that is unintelligible to the attacker. The attacker may successfully get the message, but it is unreadable. The encoded information needs to be decoded first by using the key set by the sender. The sender and receiver use the same key to convert text into cypher text and then the cypher text back to plain text. Moreover, these techniques help to scramble the data, but it can still be breached by hackers as per vulnerabilities of each method hence Steganography is a valuable method to apply as it hides the data, and it becomes invisible [35].

This research proposes a novel multi-approach steganography detection method in the context of audio files and considers the existing steganography and cryptography techniques. This work also elaborated on the current state of the art in terms of proposed hidden evidence detection and extraction approaches, the current challenges and limitations in the field of audio forensics as well as future research directions. This paper is composed of five sections, in section 2 we introduced a comprehensive literature review with a deep critical analysis of related works. In section 3, we detailed the proposed research methodology and reflected on the choices made. In Section 4 we discussed the proposed Multi-Approaches Steganographic Content Detection and Extraction providing a detailed diagram. In section 5, we presented the implementation and testing of the system we named StegoHound which embeds the proposed Multi-Approaches method and presented the obtained results with a comprehensive analysis. In section 6, we discussed the obtained results and provided a conclusion and further research directions.

## 2. Literature Review and Related Works

The current advances in the use of cryptographic and steganographic techniques to conceal information within audio files, methods and techniques used to automate the process of extracting that information are our starting point to conquer an improved approach towards an efficient automation of steganography detection and extraction. Dhamija and Dhaka [4] introduced a steganography technique in which the actual information is hidden in another object such as a sound, image, video, or watermark, however, because improved strategies emerged concurrently with this prominent level of attacking strategies, the attackers also used them. Steganography in the video was once thought to be one of the best ways to achieve high security where the sender conceals the data in the video's cover image or uses the LSB technique [6]. Before the development of video-based steganographic techniques, audio-based steganographic approaches based on spread spectrum encryption were widely used, which were later improved by employing the Direct Sequence Spread Spectrum (DSSS) phenomenon [8] as the Least Significant Bits became overused. Moreover, clustering is also considered a popular and emerging technology in the real world. Recently, a K-means clustering algorithm was introduced that outlines an improved way to hide information in an image. The pixel of the image retrieved from the video is chosen to hide the information using the K-means clustering mechanism [1]. Despite these advanced techniques, hackers still found some loopholes to attack and access information. Currently, several researchers are working to develop more useful and robust approaches that will be difficult to break. For this reason, many implementations based on combined and hybrid mechanisms are already advancing and encryption is performed with audio steganography using the DWT strategy [3]. Another attempt was made to secure the transmitted messages by using a combined strategy based on DES for DCT with image steganography was introduced in [7]. In addition, the steganography approach employing the Fibonacci sequence [5] conceals the data using a traditional flow but employs the method to generate the values to encode the pixels utilising the Fibonacci sequence concept is innovative and provides a good advance in the area. Furthermore, a multi-layered approach was recently used in which information encryption was based on genetic algorithms and residual numbers to achieve a secure environment for message transmission [2]. In Total, we enumerated eight families of detection and extraction approaches.

### 2.1. A Steganographic Approach Based on Clustering for Secure Data Communication

Secure communication is difficult these days, and many researchers have proposed various solutions to overcome this security challenge, but they still have numerous flaws. SaiKrishna et al. [1] identified a trending issue and a proposed strategy for implementing a solution to address it two to improve the security factor while transmitting messages based on LSB substitution. Other researchers contributed to the research by combining steganography with K-means clustering techniques [9].

The clustering distinguishes the article from previous studies; the authors discovered that by using the K-means clustering procedure, the pixels can be divided to perform secure communication. The proposed methodology has presented a novel approach to improving communication security, but the actual implementation of the methods was proposed as future work. Figure 1 shows the basics of the approach taken which also included an encryption phase as opposed to Steganography only.

The work in [10] articulated clear and concise facts in the steganography detection domain understanding and conducting experiments, it provided a comprehensive comparative analysis based on the output of previous works notably [11-17].

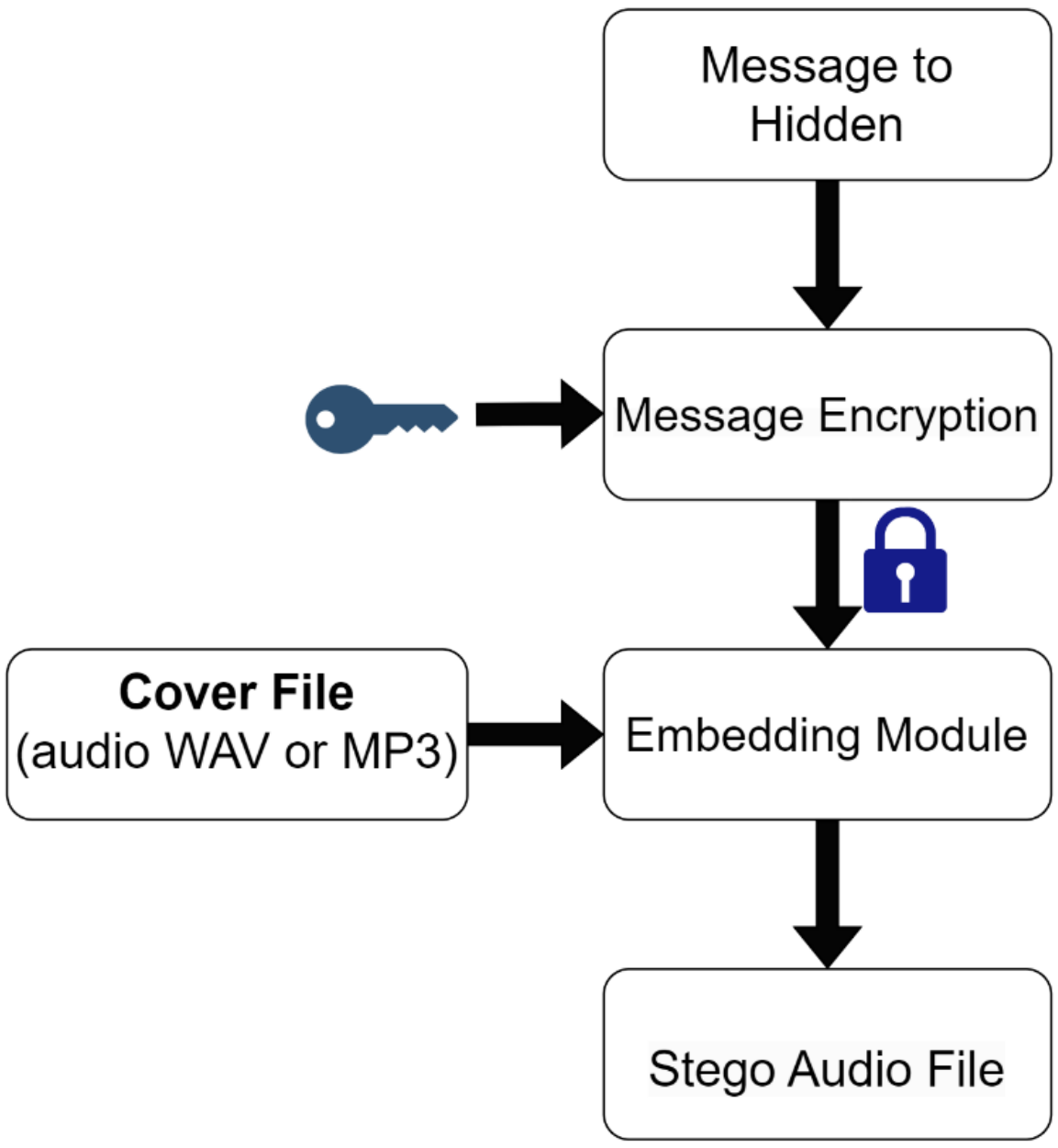

*Figure 1: Modern Steganography Information Hiding Phases in Audio Files.*

### 2.2. Encryption and Decryption of Multi-Layer Data using Genetic Algorithms and Residual Numbers

Baagyere work discusses some new techniques that are not related to text encryption or traditional image, audio, and video steganography to make digital data more reliable and secure [2]. A multi-layered data encryption approach based on the genetic algorithm (GA) as per Figure 2, and residual numbers (RN) were used for this purpose. According to the facts, GA belongs to the category of evolutionary algorithms, whereas RN is related to traditional number systems. The researchers focused their efforts on the most effective methodology for the system and implementing it for experimental research to validate where the secure data was discovered, and it provided a secure and reliable steganographic mechanism at the expense of affecting the capabilities of embedded objects.

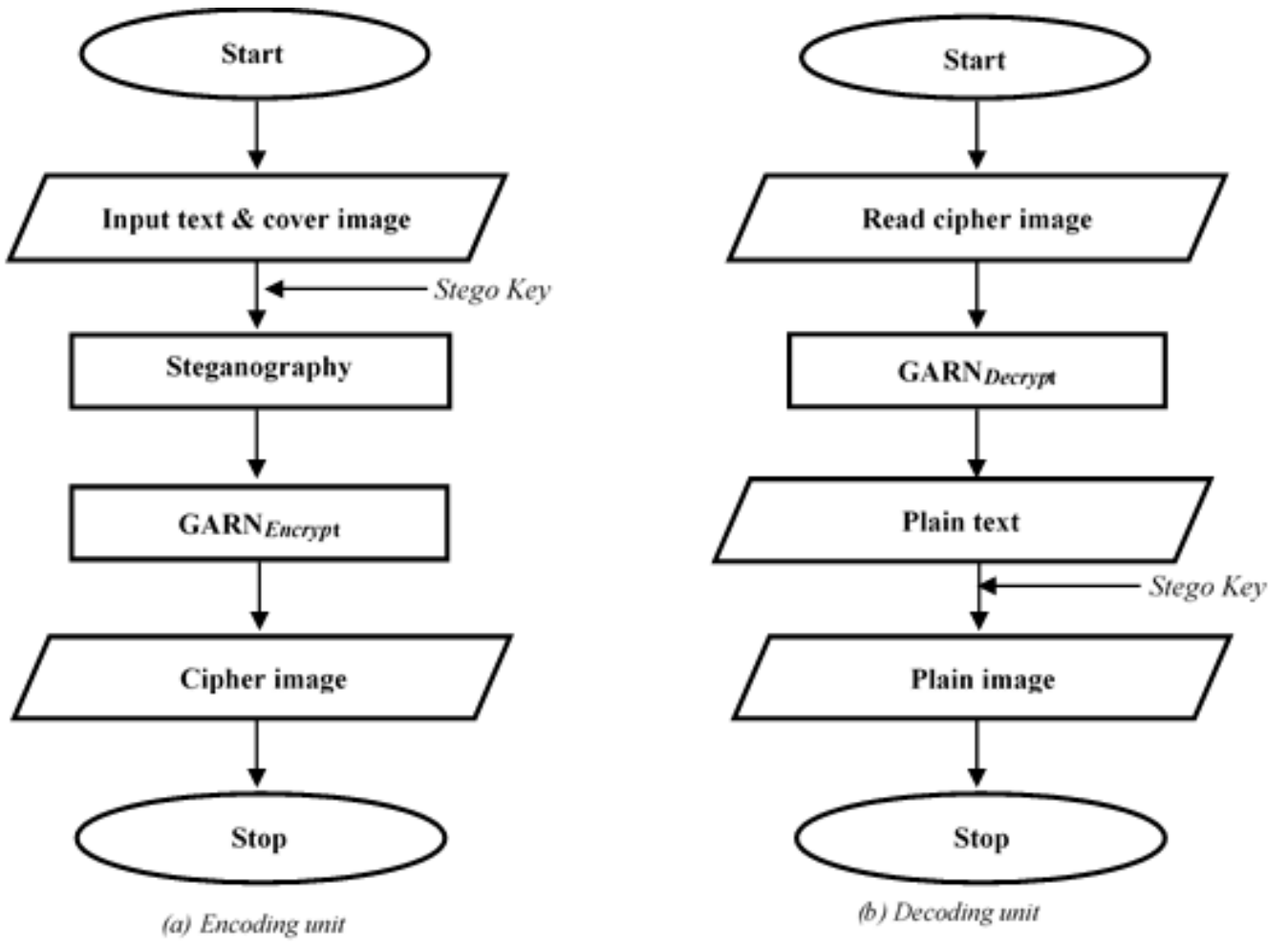

*Figure 2: GARN Algorithm Flowchart [2].*

In [14], the researchers considered the speed and compared the performance of their proposed scheme to that of existing schemes on key parameters. Complementary works [15] and [16] provided various debatable points on the proposed methodology can be modified to meet the accuracy and effectiveness needs, covering nearly every aspect of the targeted idea of image steganography only, but utilising the same phenomena for audio and video is possible as well as applying a practical implementation to test its method.

### 2.3. DWT-Based Secure Data Transfer

Another recent method of security was described by Geethavani, Prasad, and Roopa [3]. Several researchers have recently done outstanding work in the field of security to improve the privacy and confidentiality of data based on feature integration [19]. The researcher believes that the features of cryptography and audio steganography can be combined to produce a more advanced approach because the strength of one algorithm increases when it is embedded with another popular technique. It referred to a double lock approach in which the message or plain text is fed into a modified blowfish algorithm, which converts the text to cypher text.

Following that, the second phase involves performing steganography on audio and encrypting it with Discrete Wavelet Transform (DWT). The orthogonal and biorthogonal DWT were used for different purposes in the study. This approach is complicated enough because, in the DWT mechanism, the sender sends the cypher text, which is then converted to the wavelength and reserved at the receiver's end [18] as per Figure 3.

The selection of these two demanding techniques resulted in distinct and positive societal outcomes, suggesting a better idea to improve information security and channel security [20]. With the help of experimental results, the articulated facts are well-defined, and the validity of the proposed system is clear. This technique can make hacking and other attacks more difficult to carry out.

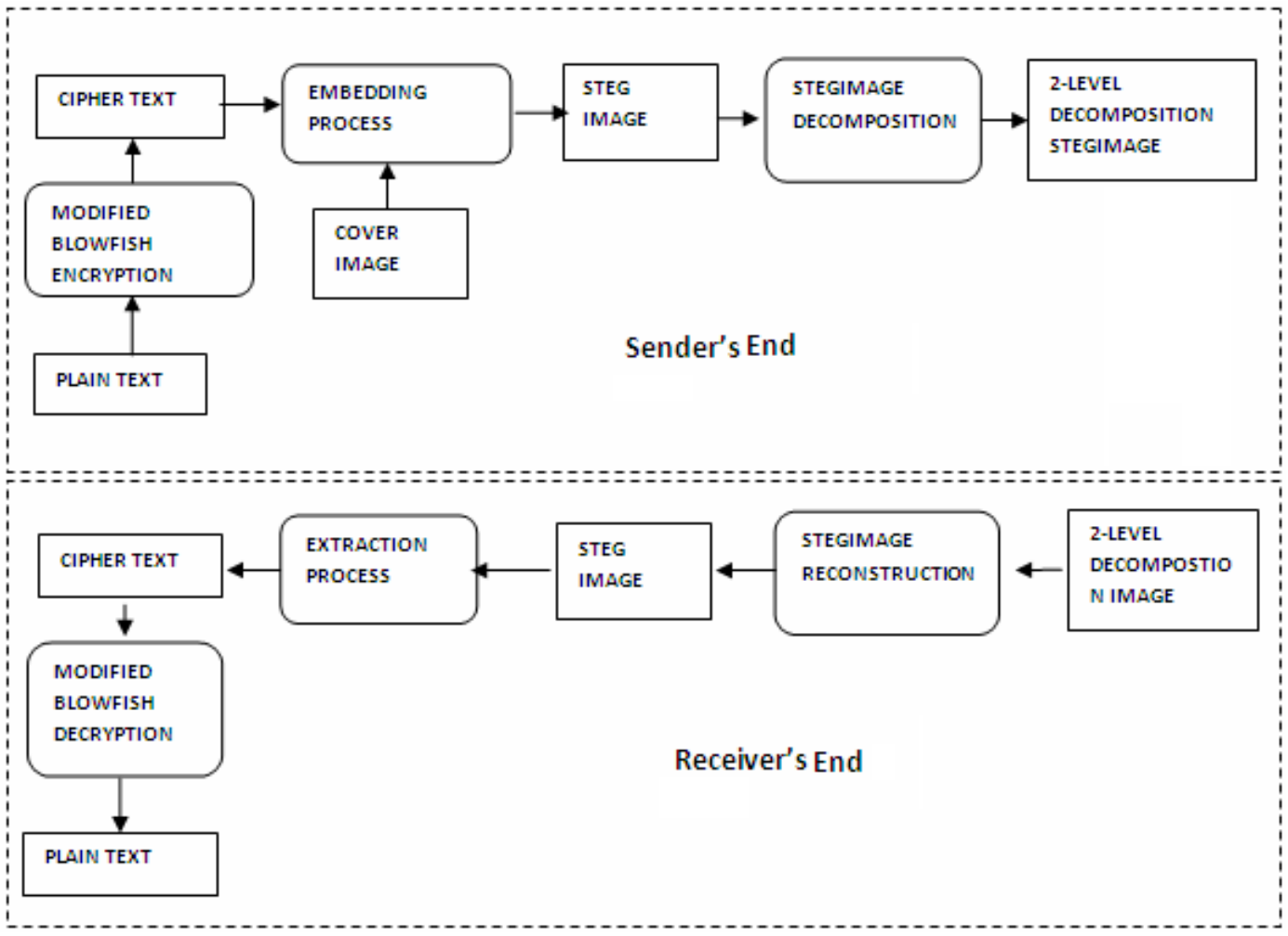

*Figure 3: DWT-based secure data transfer Framework [3].*

The uniqueness of the concept has left an open horizon for future research. However, there are some limitations to the type of steganographic approach used because one technique is ideal for a few scenarios, but it may not work in all other attacks using a different type of file as opposed to the audio file used in their experiment.

### 2.4. Steganographic and Cryptographic Method for Safe Cloud Data Migration

Despite the numerous advantages of cloud computing, it has also faced several security and privacy concerns. Dhamija and Dhaka [4] highlighted cloud computing's rapid adoption, and the author commented on the high availability and remarkable performance of this emerging technology [4]. In this article, the researchers take a stance on security issues and implement a composite strategy to provide high-level security to data transferred over the cloud. The researchers used a two-way security mechanism based on a combined encryption algorithm and steganography technique [13]. Encryption was previously used to secure messages, but this technique was sometimes replaced by improved approaches of steganography which all had their vulnerabilities. As a result, the authors proposed SCMACS: Secure Cloud Migration Architecture Using Cryptography and Steganography, a novel methodology to conceal data within images [17] as per Figure 4.

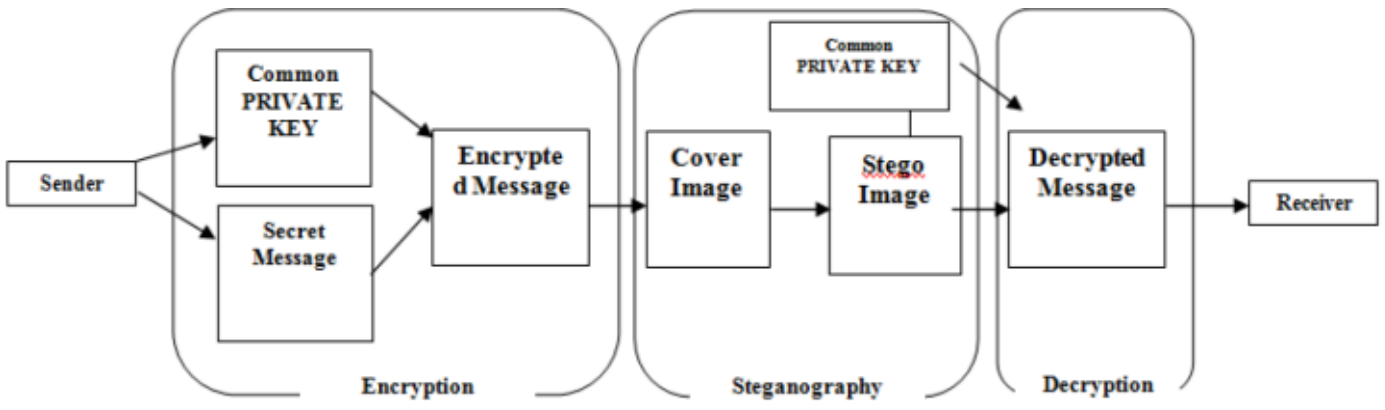

*Figure 4: Steganographic and Cryptographic Process [4].*

This provides secure data transmission to databases, later being transferred to the receiver in encrypted form by using the proposed approach to secure the information and reduce the likelihood of data loss. The dynamic values generated for the private key make the strategy more difficult to access and it needs further work on implementation to perform more tests to reinforce the concept using current trends such as video cryptography as opposed to image steganography as it is considered more secure in terms of making steganalysis more difficult.

### 2.5. Steganography with Sub-Fibonacci Sequence

Security breach attacks are carried out each year to obtain useful information from individuals for various purposes. According to Aroukatos et al. [5], concealing information securely is an arduous task as his team was working on an approach where the data is hidden using a traditional flow with the values to encode the pixels generated in a novel way utilising the Fibonacci sequence concept. The Fibonacci sequence is used to generate the key, which improves on the previous technique [20] they worked on using the LSB of RGB technique. The article clearly explained the LSB approach's workings and limitations, which were later covered by using the Fibonacci sequence which produced results by using a base system other than binary with an approach which has not been exhausted by previous researchers [5]. The Fibonacci sequence provides an improved mechanism for making data more secure and safe than the LSB approach. The key is difficult to gather by criminals due to the complexity of this sequence. A steganalysis approach cannot decode the sequence easily. More research on this technology is recommended to determine the detailed criteria for selecting the encoding sequence.

### 2.6. DNA and KAMLA Approaches in Metamorphic Cryptography

According to Singh et al. [6], life today would be deemed incomplete without the Internet because everyone has something to send and something to receive online, hence, data secrecy and privacy are crucial. Data transit over the internet was previously secure using encryption mechanisms. Several cryptographic schemes were proposed to improve security and the current methods at the time, including the RSA, the RSA method based on a singular cubic curve, JCE, a metamorphic cryptography method, an encryption method based on an indexing chaotic sequence, ICSECV, a real-time video encryption method, and the metamorphic cryptography method using auditory information via a carrier wave. [21]. These cryptographic schemes quickly became outdated for advancing cyberattacks and the availability of powerful resources for most. As a result, attackers can launch a brute-force attack on encrypted data transmitted over the internet.

The DNA and Kamla method addresses the potential issues in securing internet-based information transmission. The technique uses a DNA approach to encrypting data transmitted over the internet using a seven-step process that includes some special steps at the sender's end such as after converting the data to ASCII bits, a DNA process is applied to the resulting keys and data, and the KALMA technique is used in the final step to hide the message in DNA bases [22-25]. To decrypt the received encrypted message at the receiver's end, the reverse approach is used to obtain the original message. This proposed scheme outperforms previously used techniques in terms of robustness, payload size, visual detection, and steganography in this context. Combining cryptographic techniques such as steganography methods and encryption resulted in a highly secure environment, as the DNA approach resulted in satisfactory robustness and the use of the KALMA approach resulted in a high payload. To expand on this work, different carrier mediums and DNA versions can be used to improve the payload and security of digital data transmission [6].

### 2.7. DES and DCT-based Steganography

The frequency of attacks on confidential data rose as technology improved [23]. The methods of cryptographic methods of encryption and steganography are two of the most widely used techniques for protecting digital information, but several researchers have pointed out security concerns and suggested a reliable way to safeguard the data. According to Ramaiya [27-28], the best approach is to combine methodologies and they proposed two popular approaches, incorporating the DES algorithm and another via the discrete cosine transform (DCT) to overcome the vulnerabilities of older approaches. Solichin and Ramadhan [35-38] proposed a method for using the application to secure Word, Excel, PDF, and PowerPoint documents into their secured versions [36]. Due to the high demand for DES, the researchers prefer to use that algorithm for encryption and then use the DCT approach to reveal the message hidden in the cover image, again combining different techniques that were already in use to strengthen security. The combination of DES and DCT steganographic methods proved to improve data security but to achieve a high functional rate, the computational time must also be reduced. The researchers conducted the experiments to provide evidence for the proposed approach, and the results were as expected [8-10] widening the previous research works on the steganographic use of combining various cryptographic methods such as RSA, DES, AES, and Blowfish, but they have not used DWT comparison in this instance [39].

### 2.8. Encrypted Audio Steganographic System Based on Spread Spectrum

According to Anjana et al. [8], communication always requires a smooth and secure path. Several algorithms have been introduced to provide a secure channel to transmit data, but some have failed due to their poor tackling power [24].

Researchers have used audio steganography to improve the security mechanism [29] based on the Direct Sequence Spread Spectrum (DSSS) scheme, and by combining steganography and encryption, the algorithm provides additional security [26]. BER, Ratio of Signal to Noise, Peak Signal to Noise, Vigour Feature, Chip Frequency, and other performance measures were used to evaluate the proposed system's imperceptibility, security, detectability, distortion, capacity, and accuracy, providing a secure and reliable steganographic mechanism at the expense of affecting the entrenching volume [35]. Figure 5 shows the embedding process of this technique. This DSSS mechanism is a significant step toward improving traditional methods providing a novel direction of work as well as computational metrics that are useful for implementation and decision-making due to their measured values.

The system is developed to hide messages such as cyber intelligence text, where small messages must be communicated, and it is associated with elevated levels of required security. However, increasing the capabilities to cover audio content could improve the message steganographic capacity limitation, which is due to a wide range of parameters used in the experimental designs and numerous other factors to obtain more improved results. A change in the factors chosen could have a wider impact on the contribution [8] of the research.

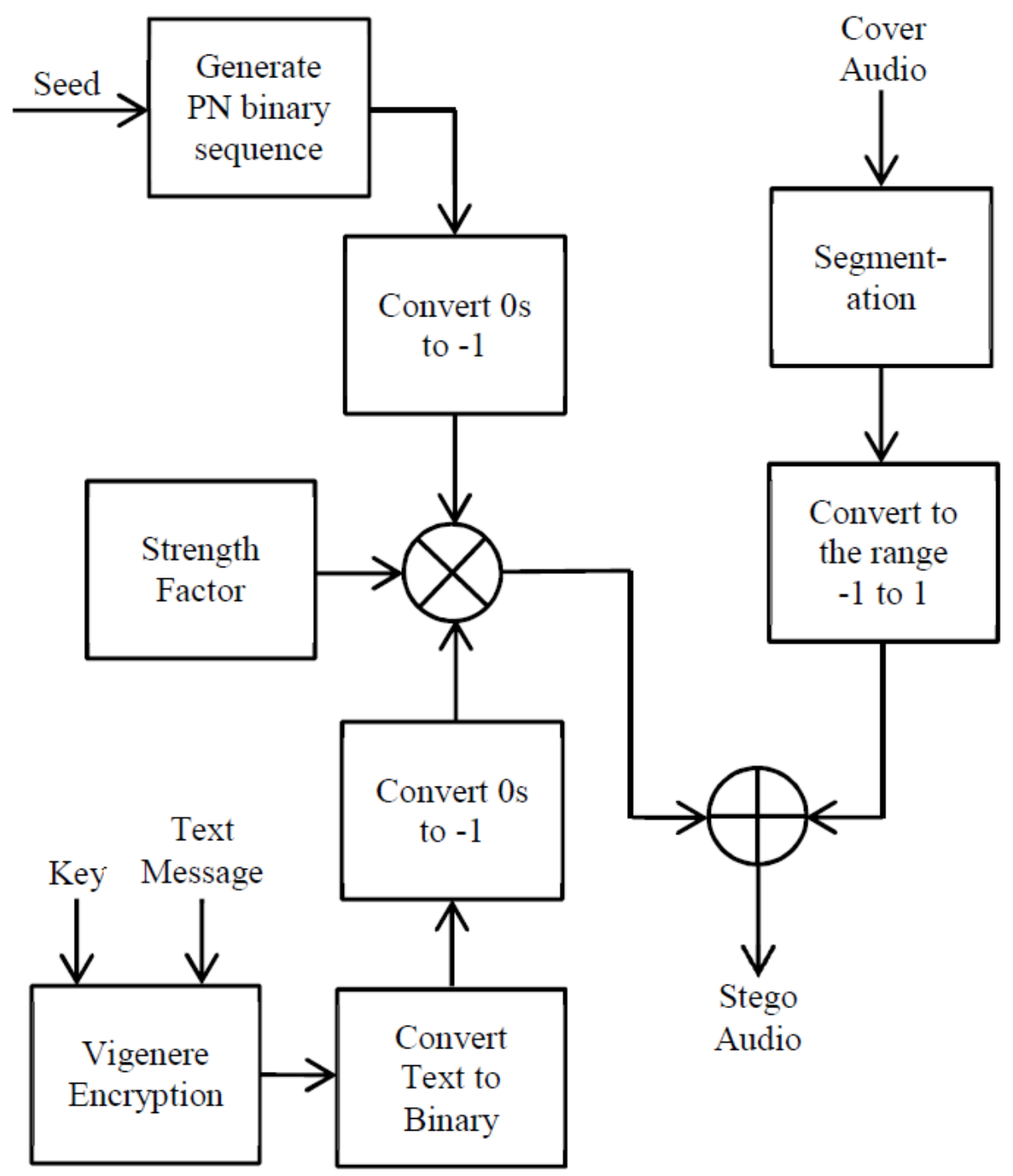

*Figure 5: The Standard Embedding Process [8]*

## 3. Research Methodology

The methodology for this work is a blend of conceptualisation, development and experimentation which was explicitly designated to problem-solve the issue of automating the extracting of evidence hidden with cryptographic and steganographic techniques, the output of this work is a multi-approaches system that ensures an efficient, effective and accurate automated detection of steganographic and cryptographic artefacts audio files. In this research, the chosen methodology is the following:

1. Achieving an in-depth domain understanding through literature review and critical analysis of the related works.
2. Shortlisting, selecting and validating appropriate steganographic and cryptographic detection methods.
3. Determining the sequential order of the selected methods in the context of experimental conditions.
4. Introducing scheduled modifications to one or more variables.

5. Testing the proposed multi-approach detection in different size WAV and MP3 files and measuring and controlling the used variables.

### 3.1. Decision on sample datasets

The most serious perceived issue with steganography from a forensics point of view is never being 100% assured of the cleanness of the datasets being free from hidden artefacts and the type of files that host data can accommodate. The most common form of steganography uses a file type that is not commonly used. Image steganography can sometimes be seen as the first option and exhausted; thus, audio files will be the sample data set. WAV files, in particular, that contain encrypted data, such as a zip file, are preferred for this experiment. We designed and implemented a dataset generator which allowed us to generate 320 different lengths of audio files varying from 10 seconds to 1600 seconds and with a proportionate number of steganography contents hidden in each one. This dataset is divided equally between WAV and MP3 audio files.

### 3.2. Data Analysis Models

This research uses two statistical models such as inferential statistics to allow for the prediction of a hypothesis based on the data obtained from testing which was tampered with to analyse it as a potential steganographic file. Many unprocessed artefacts are the result of the data-hiding technique used in this research. Moreover, the data obtained from the testing tools were analysed using empirical data analysis.

### 3.3. Result Analysis Metrics

The results were analysed by testing between current tools and the project toolkit to see which tools successfully identified steganography, which tools detected encryption, and which tools could identify and uncover the sample data. The tools were tested against a variety of sample data sets, including WAV and MP3 files and Encrypted text and graphic files. The hidden data used for testing were either ZIP files, Encrypted TXT, PNG or DOCX Files. To eliminate discrepancies, each tool was tested multiple times and with the same type of data. To sum up, the metrics adopted for evaluating the efficiency and accuracy of the automation are the detection rate in the form of the number of hidden artefacts identified and extracted as well as the number of false positives and false negatives.

## 4. Proposed Multi-Approaches Steganographic Content Detection and Extraction

In this section, a detailed account of the proposed multi-approaches detection mechanism is provided. Initially, the proposed method accounts for any captured data whether stored or transmitted in a digital format and thus can be used as evidence in a legal case and is referred to as digital evidence. This includes Emails, text messages, media content, social media posts, computer files, and other types of digital data. The following are the twelve phases constituting the multi-approaches method proposed.

### 4.1. Digital Evidence Acquisition

The process of obtaining digital images for forensic analysis is referred to as image acquisition. Capturing images from a variety of sources, such as digital cameras, smartphones, tablets, and other digital devices, falls under this category. However, the image acquisition process is important in forensic analysis because it ensures that the original image is preserved in its original state to preserve the evidence's integrity.

### 4.2. Live Forensics Analysis (automated)

The process of gathering and analysing digital evidence from an active, or live, computer system is known as a live forensics examination. This may entail performing tasks like scanning the RAM of a computer, analysing currently active processes, and logging network traffic. Contrary to conventional forensic examination, which frequently entails taking a snapshot of a computer's hard drive and analysing the data offline, this procedure involves no such steps.

### 4.3. Forensics processing, indexation, and deleted file reconstruction

The process of analysing and deciphering digital evidence gathered during a forensic examination is known as forensic processing. However, this may also entail activities like locating and extracting pertinent data, confirming the legitimacy of the supporting documentation, and maintaining the accuracy of the data throughout the analysis process. Deleted file reconstruction is the process of recovering deleted files from a digital device, and this can be done using specialised software to recover the deleted data and then attempt to reconstruct the original files. Indexation, on the other hand, is the process of creating an index or catalogue of the digital evidence that has been gathered for additional forensics analysis.

### 4.4. Indexed and Recovered WAV and MP3 Files

Audio files that have been located and restored from a storage device or backup are referred to as indexed and recovered audio files. However, the files may be in the digital audio file formats WAV or MP3, respectively. Nevertheless, MP3 files are smaller in size and have lower audio quality, while WAV files are typically larger and have higher audio quality. These files can be recovered using specialised software even though they may have been deleted, formatted, or lost due to other types of data loss.

### 4.5. Tool-automated steganography detection

Software programs that can automatically scan digital files and detect the presence of hidden data are known as automated steganography detection tools. These tools operate by analysing digital files for patterns or anomalies that may indicate the presence of hidden data. However, statistical analysis of file headers, footers, and other metadata are some common methods used by automated steganography detection tools in identifying hidden data.

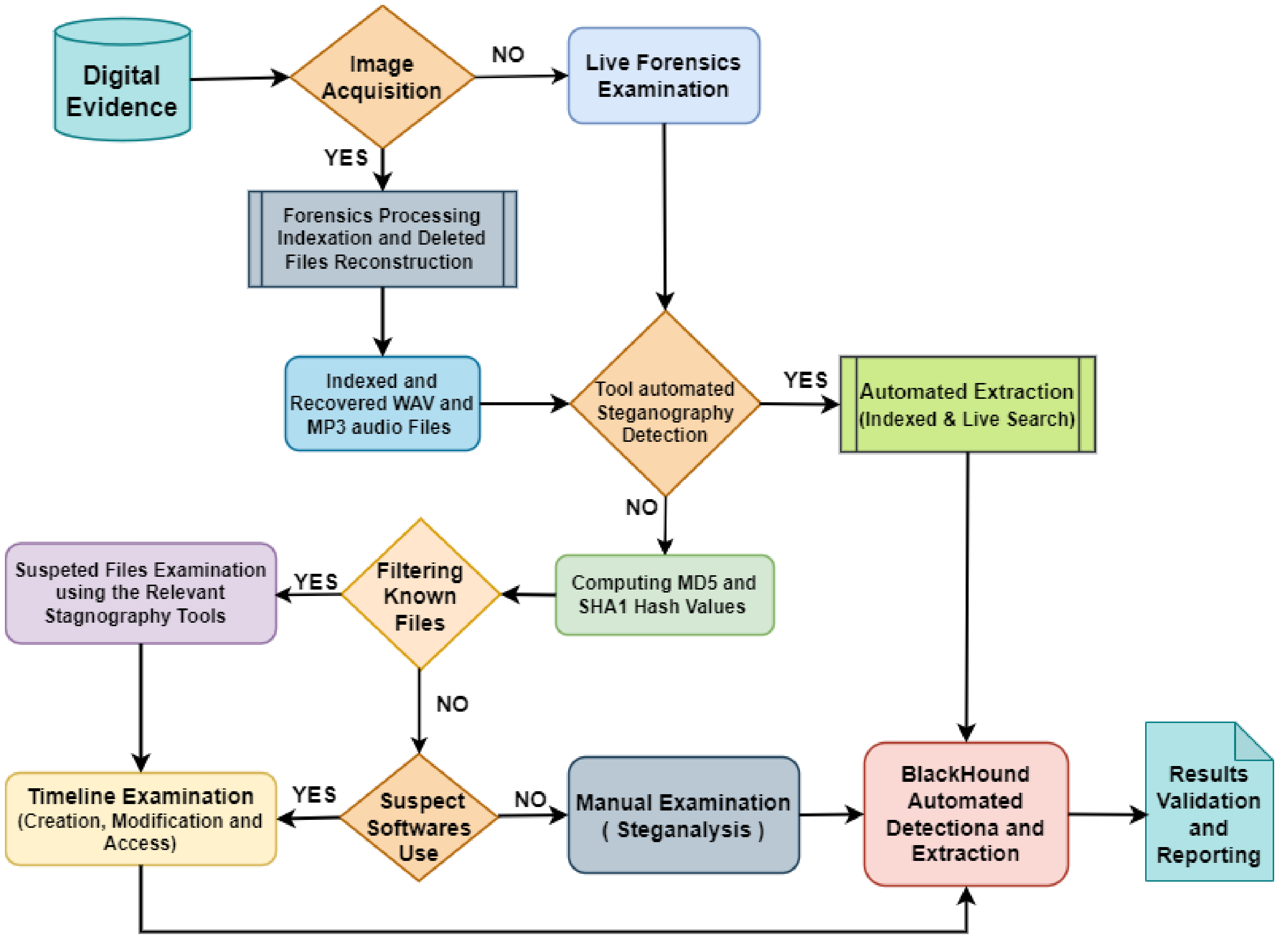

*Figure 6: Proposed Multi-approach Setagnography Detection and Validation Algorithm.*

## 4.6. Flagged files (Known Files Content)

The process of identifying and marking specific files or pieces of data within a set of digital evidence for further forensic analysis is referred to as flagged files or indexed searches, which can be done manually by a human forensic expert or using specialised software. The human forensic expert, on the other hand, may use a keyword search to identify such files relevant to the case and then flag or index those files for further review. When performing such tasks, however, considerable resources are required.

## 4.7. Suspected Files Examination using the Relevant Steganography Tool

Examining suspect files with the appropriate steganography tool is the process of detecting and analysing hidden data within digital files suspected of containing steganography. A forensics human expert may use a variety of techniques to identify suspect files during a forensic examination, such as keyword searches, file hash comparison, or statistical analysis of file headers and footers. Once suspect files have been identified, the forensics expert can analyse them with a steganography tool to detect the presence of hidden data.

## 4.8. Filtering using Known File Format KFF

Filtering known files refers to the technique used by a forensic human expert to identify and remove files from the collected set of digital evidence that is known to be irrelevant. This can be accomplished by comparing the files in the evidence to a database of known files, such as operating system files, common software files, and other types of files that are unlikely to be relevant to the case. However, this file filtering technique can help reduce the amount of data that must be manually reviewed by a human forensic expert.

## 4.9. Computing MD5 and SHA-1 Hash Values

The MD5 algorithm generates a hash value of 128 bits, while the SHA-1 algorithm generates a hash value of 160 bits. Both algorithms are considered one-way functions. However, in digital forensics, hash values are used to identify and track specific files, verify the integrity of files collected as evidence, and detect if a file has been tampered with. A human forensics expert can manually compute the hash or rely on specialised software such as FTK, AXIOM or UFED to determine if two files are identical by comparing their hash values.

### 4.10. Timeline Analysis (Creation, Modification, and Access)

The process of creating a chronological representation of events that occurred on a computer or digital device is known as timeline analysis. This may include file creation, modification, and access times, as well as other system events like logon and logoff times and network connections that may be stored in RAM. Timeline analysis, on the other hand, is a significant practice in digital forensics for identifying patterns of activity or reconstructing events.

### 4.11. Suspect Software Use

The identification and analysis of software that is suspected of being used in cybercrime or other legal issues are referred to as suspected software use [36]. Forensic human experts may examine the contents of a computer's hard drive to identify specific software programs that have been installed or network traffic may be examined to identify the use of specific software tools or protocols to collect traffic data without the administrator's knowledge.

### 4.12. Manual Indexed Forensics Search

A manual examination live search is a process in which a human forensics expert examines a computer or other digital device in real-time while looking for specific files or information. This method can be used for a variety of purposes, including conducting forensic investigations, locating evidence for a legal case, and recovering lost or deleted data. During a digital forensics investigation, the human forensics expert searching will typically use specialised software and techniques to search the device. Nonetheless, this live manual examination search is performed in real-time, allowing the human forensic expert to quickly retrieve information and make informed decisions. the proposed solution functioning is graphically summarised in Figure 6.

## 5. Design, Implementation and Testing

The proposed multi-approaches concatenate in a precise order common detection and analysis techniques to deal with steganographically hidden content. Therefore, detecting steganography in MP3 and WAV files may require a combination of different techniques and tools to increase the likelihood of detection. In the first stage, the StegoHound algorithm uses the Statistical Analysis function (SAF) to identify any subtle changes in the statistical properties of the audio files and look to identify deviations from normal patterns. The output of the first stage is crucial to deciding the next stage, if the SAF output is negative meaning no statistical abnormalities were detected, StegoHound will proceed with Spectrogram Analysis to detect any steganography track through the observation of modified frequency components of an audio signal. Analysing the spectrogram may reveal hidden patterns or anomalies and StegoHound uses Spek to generate and analyse spectrograms. In the third stage, StegoHound operates a comprehensive File Signature Analysis (FSA) by comparing the file signatures of suspected files against known steganography tools, it employs the least significant bit (LSB) examination to reveal hidden data within the audio files and detects modifications. Finally, the fourth stage involves a Forensics Content Analysis (FCA) looking for any alteration of the audio perceptual quality. This toolkit's goal is to provide code and materials for automating the detection of hidden data as well as the retrieval of any data items discovered and as a framework to help the forensics professional find hidden data that otherwise could be missed. StegoHound is a modular framework that houses several interchangeable Python scripts for detecting encrypted and steganographically hidden data. The toolkit is written in Python 3 and consists of three Python scripts and two test files. The full code and dependencies are provided for free on the GitHub platform.

### 5.1. *System* Architecture

The system architecture of StegoHound is based on a standard module structure. Modules are used to integrate all the toolkit's functionality into separate scripts and units as per Figure 7. To verify file integrity, the source files must be loaded into the extra/music folder to create a comparable hash database.

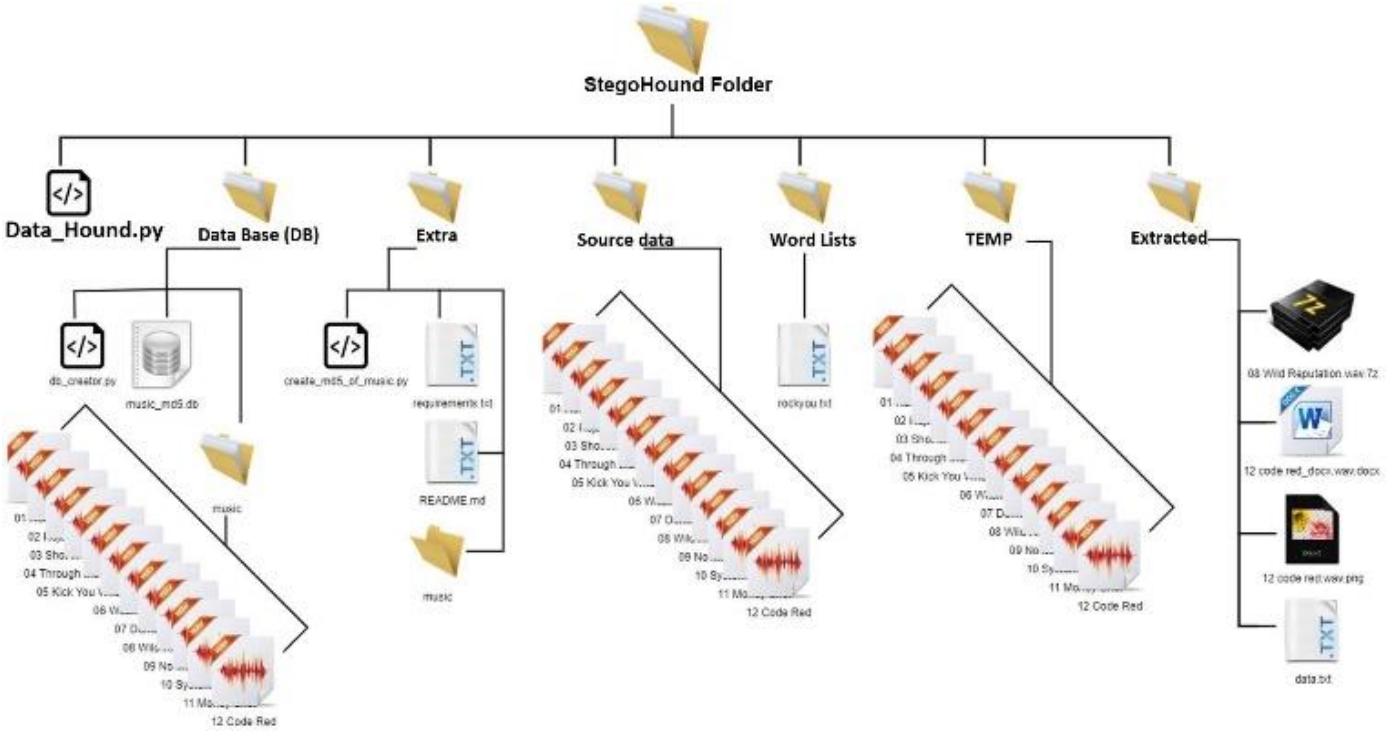

*Figure 7: StegoHound Toolkit Structure*

The create_DH_of_audioc.py script allows for manual digital hash computation; it is the shortest of the three scripts, consisting of 11 lines of code, and serves only to create a hash dictionary of the audio source files. StegoHound Toolkit maintains a database hash digest of other file types can be easily accomplished by simply changing the originating data folder and the hash dictionary output. The next step is to execute the db_creator.py script.

This script creates an SQLite3 database containing the SHA1 and MD5 hash digests of the source audio files, which can later be compared. It is currently a remarkably simple database for testing purposes with one table and the three columns ID, song name, and song hash. Different tables per album could be created and stored in a single database, depending on the number of audio files to be checked.

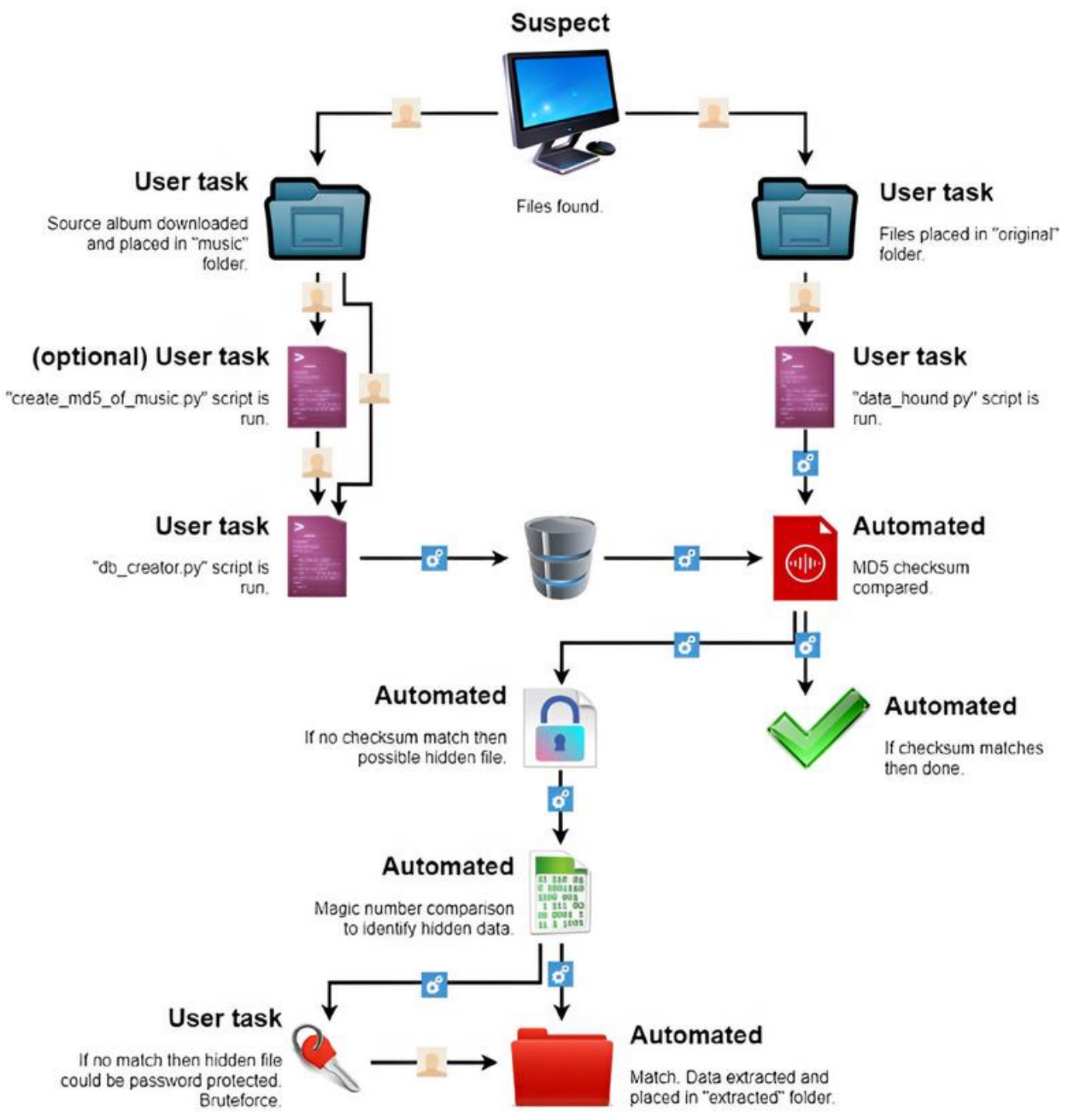

*Figure 8: StegoHound Use Case Diagram*

Due to the time-constrained nature of the investigation requiring speed and the checksums created being used only for file integrity, MD5 and SHA1 were chosen over SHA-256. The User executes StegoHound through all (batch) or individual audio files exported from the forensics image using the FTK imager. The audio files to be analysed for hidden data are saved in the /original folder. StegoHound makes copies of these files and saves them to the /original_copy, and this is for analysis, not the original. The copied files are hashed and compared to the hash of the original files in the database; if there is a difference, StegoHound notifies the user—this is the first sign of file tampering, then uses the file signature to scan the audio file for any hidden files (magic number). An overview of this process can be found in Figure 8.

Finally, after the detection process, the type of hidden file is identified, and any hidden data is extracted and placed in the extracted folder. If an audio file contains an encrypted zip file, StegoHound will, if prompted, perform a brute-force attack on the password before extracting the decrypted file
and placing it in the extracted folder. An overview of the automated methods is represented in Figure 9.

## 5.2. StegoHound Scalability

The toolkit is built in a modular structure, meaning expansion is possible with the addition of various file types, steganography methods, and encryption methods, all of which are insertable. File types are checked against file signatures, which allow identification of the file type even if the extension has been changed or is non-existent. An example of this is a.wav file that has been compared to the checksum of the source file and found to be incompatible.

This is the first indication that there is hidden content, and just underneath that, it can be observed that the magic number of a file type is being looped through the copied file, which reveals the file to contain a 7-Zip archive. The current list of file signatures to check only features the most common file types. The final check of the if-else loop is when an unknown file type is presented to alert for its existence and No match comments.

This, like many functions within the toolkit, is expandable, as more file types and their identifying file signatures can be added. In summary, the toolkit focuses on audio files, more specifically .wav files, but the toolkit is expandable in the form of additional units to focus on other file types. Still, it should be noted that various file types can currently be extracted and decrypted.

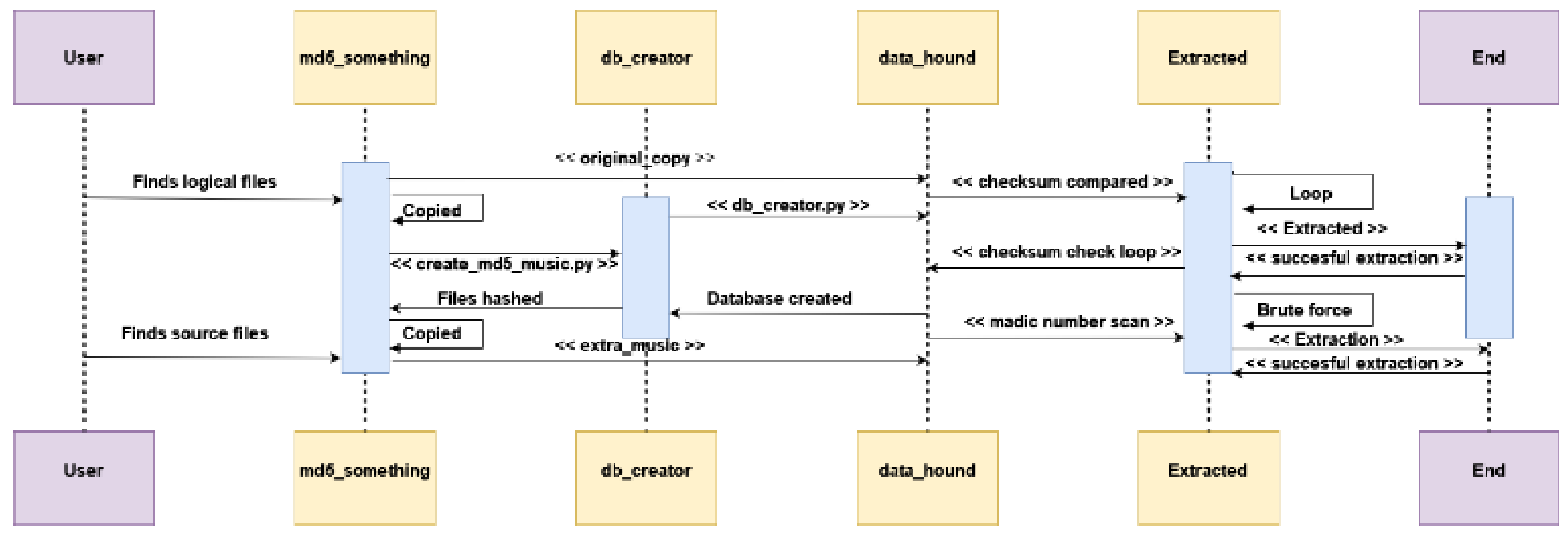

*Figure 9: The StegoHound Detailed Functioning Diagram.*

### 5.3. Testing and Validation.

Detecting steganography in MP3 and WAV files involves analyzing the audio data to identify any hidden information or alterations made through steganographic techniques in this research, we simulated the masking of docx files, text files, PNG images, and zip files as hidden messages in different size WAV and MP3 audio recordings to evaluate the efficiency and effectiveness of the detection and extraction automation. These two file extensions were chosen for this research as they are widely used and often picked by criminals to avoid attracting attention when transmitting information [20]. It was critical to test the sample data on current tools to allow a fair comparison between the StegoHound and current use cases to effectively create a toolkit to remediate the issues previously discussed within the field of digital crime as well as its vulnerabilities and weaknesses of current techniques. It was decided to put one open-source solution and one proprietary tool, both of which are commonly used in law enforcement, to the test. The open-source solution was Autopsy, and the proprietary solution was FTK. Because of the need to use the tools correctly, it was also decided that the tools would be ones that the tester is certified to use and has professional knowledge to carry out the tests efficiently. These Tools were simply chosen as comparisons due to their popularity and the tester's permission (full licence) to use them. FTK (Forensic Tool Kit) claims to automatically extract data from PKZIP, WinZip, WinRAR, GZIP, and TAR compressed files [15] and to locate all evidence with 50% thoroughness [2]. In this experiment, FTK could not detect the presence of hidden data, encrypted data, or ZIP files when testing the sample data, so no data hidden within the audio files was extracted using the FTK tool as shown in Figure 10.

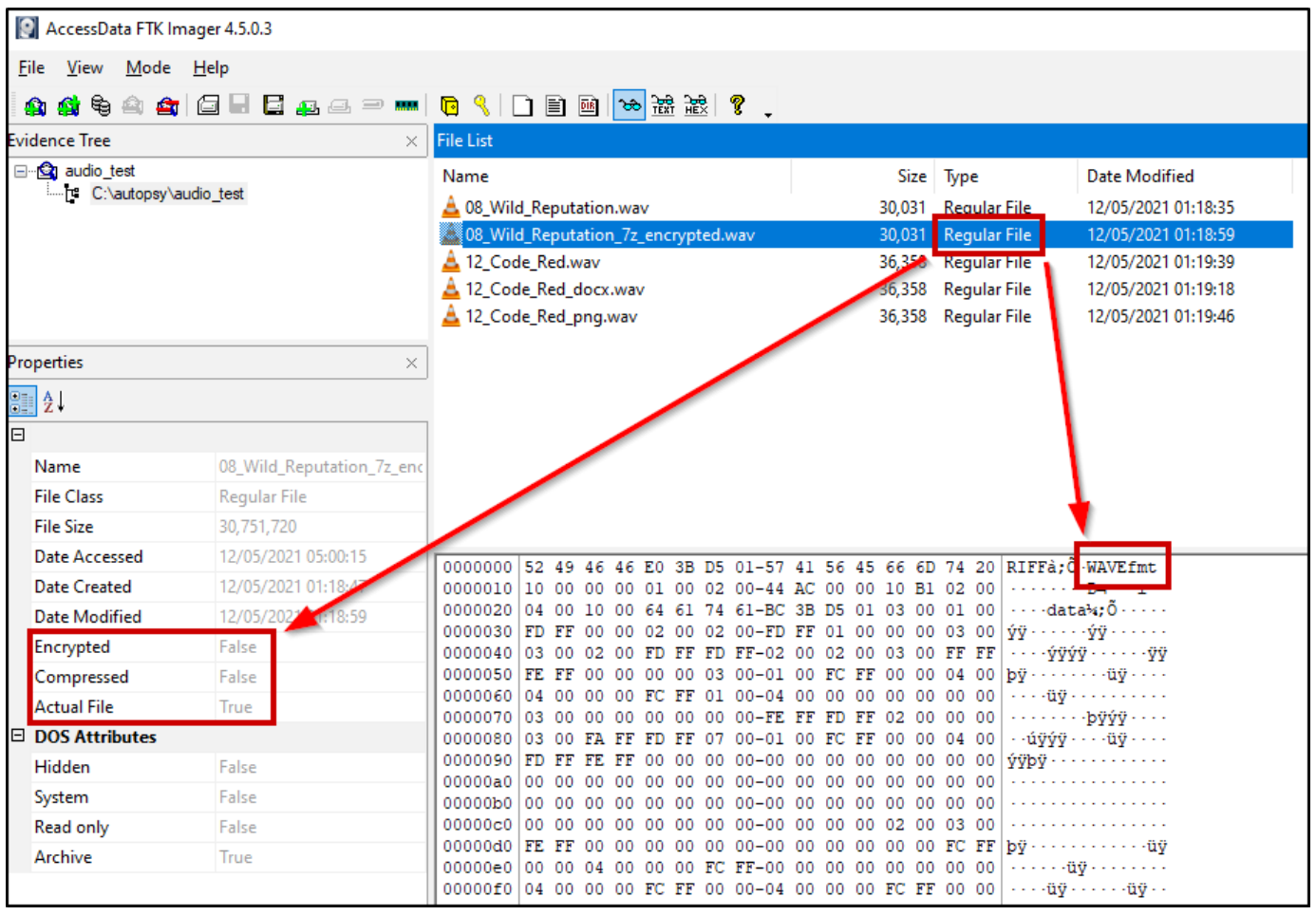

*Figure 10: Testing Sample Data in Exterro FTK*

The encrypted text file embedded within this WAV file is not indexed by FTK and AXIOM, with a minor investigative lead raised by FTK which identifies a deleted file with an unknown format flagged as a potentially encrypted file within the set of recovered deleted files. Additionally, metadata analysis and notably the Modification, Access, and Creation time data (MAC (Modification, Access, and Creation) (Modification, Access, and Creation)) illustrated inconsistencies in the file changes and creation dates and times which are often considered as an investigative lead for steganography in the Windows OSs [27]. However, similarly to FTK, Autopsy correctly found the files as audio files, but it did not find any of the hidden data within them. See Figures 11 and 12 for the number of hidden contents identified in the experiment. For comparison and consistency purposes, we carried out a manual hexadecimal investigation using WinHEX and We have carried out manual analysis following expert forensics examiner recommendations.

The obtained results confirm the fact that StegoHound outperforms both manual and FTK automated in terms of the number of extracted data masked by steganography and cryptographic techniques. The main differences in the number of data extracted are heavily affected by the audio file format WAV or MP3, the finding suggests that data hidden within MP3 files are more likely to be discovered in different size audio files with lengths varying from 50 seconds to 1500 seconds.

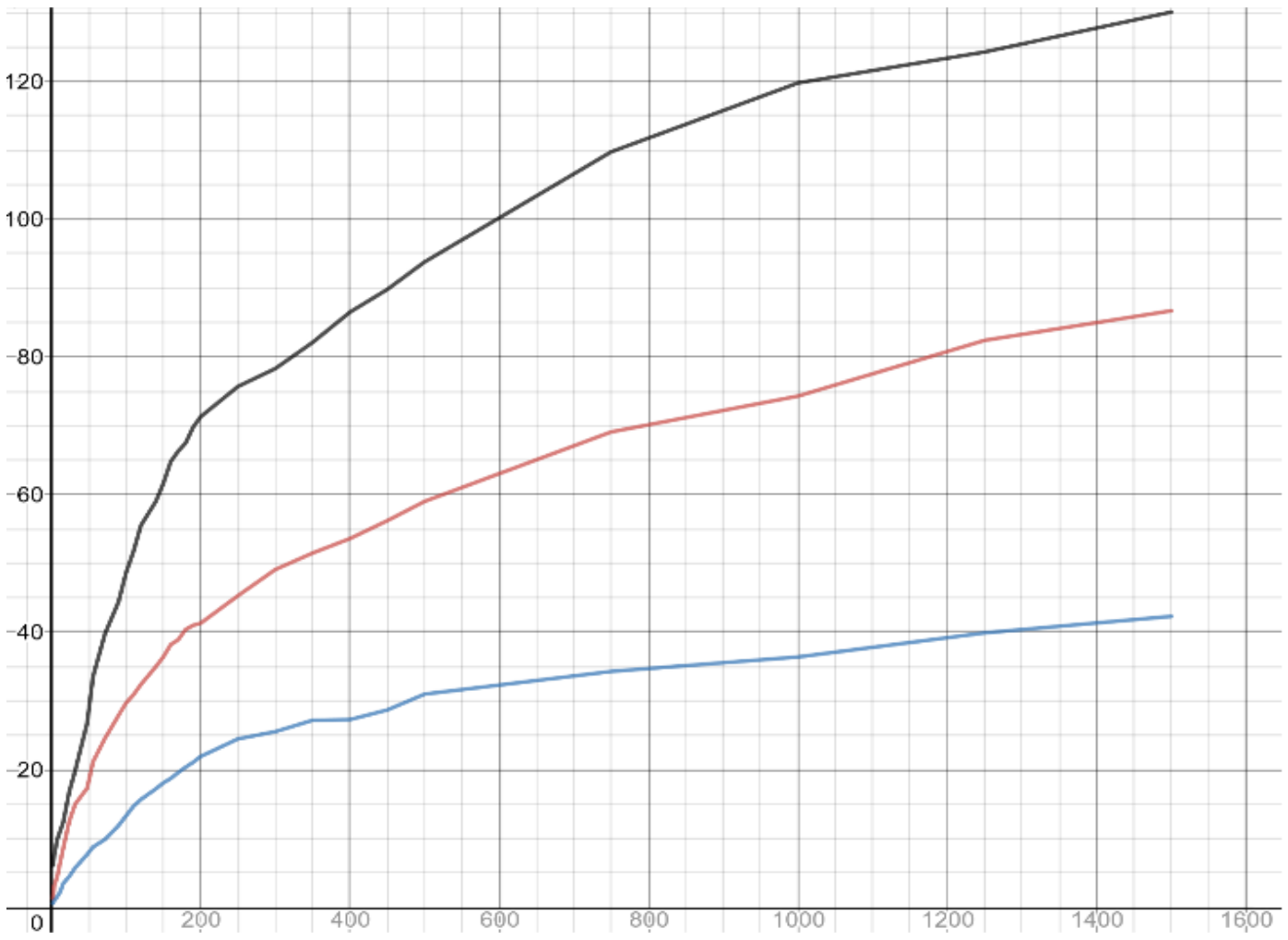

*Figure 11: Number of Hidden objects (text or graphics) detected in distinct size WAV audio files (in seconds) using StegoHound automated (black), FTK (red) and manual HEX-Search (blue)*

This is due to the compressed nature of MP3 files often smaller (compressed) compared to WAV. Nonetheless, the gap is gradually closed and the masked and hidden artefact using steganographic and cryptographic detection rates in longer audio files is almost identical.

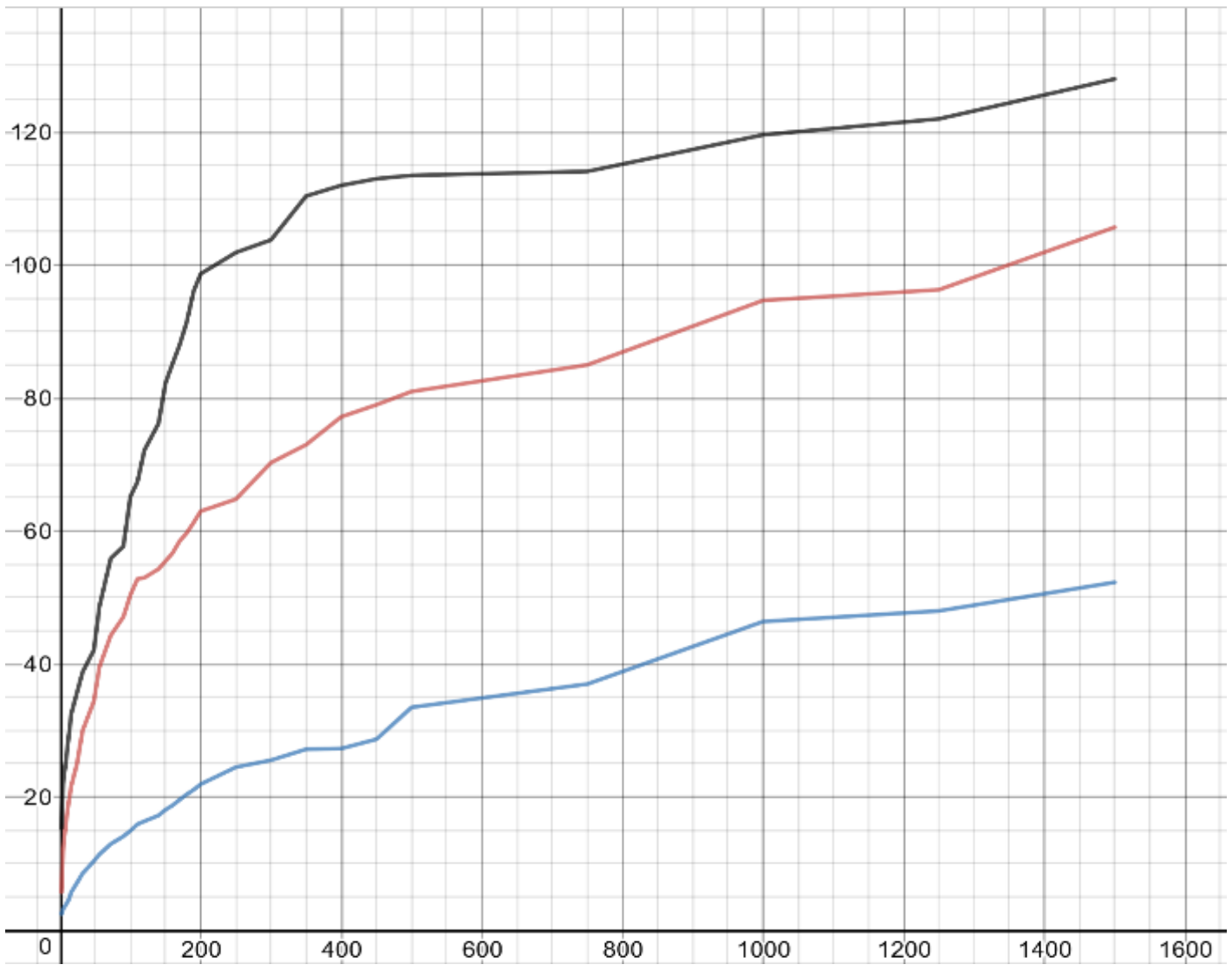

*Figure 12: Number of Hidden objects (text or graphics) detected in distinct size MP3 audio files (in seconds) using StegoHound automated (black), FTK (red) and manual HEX-Search (blue)*

In addition to these suggestions, an expansion of file signature types would be ideal to enable the scanning and extraction of additional file types, ranging from the most used files to the most obscure documents.

The last set of testing covered the accuracy of the detection and extraction through the measure of False Negatives (FN) rate related to the detection of files hidden in WAV and MP3 files, The findings suggest that WAV files produce a higher rate of false negatives but with lower probabilities, while MP3 account for a lower FN rate but with higher probabilities.

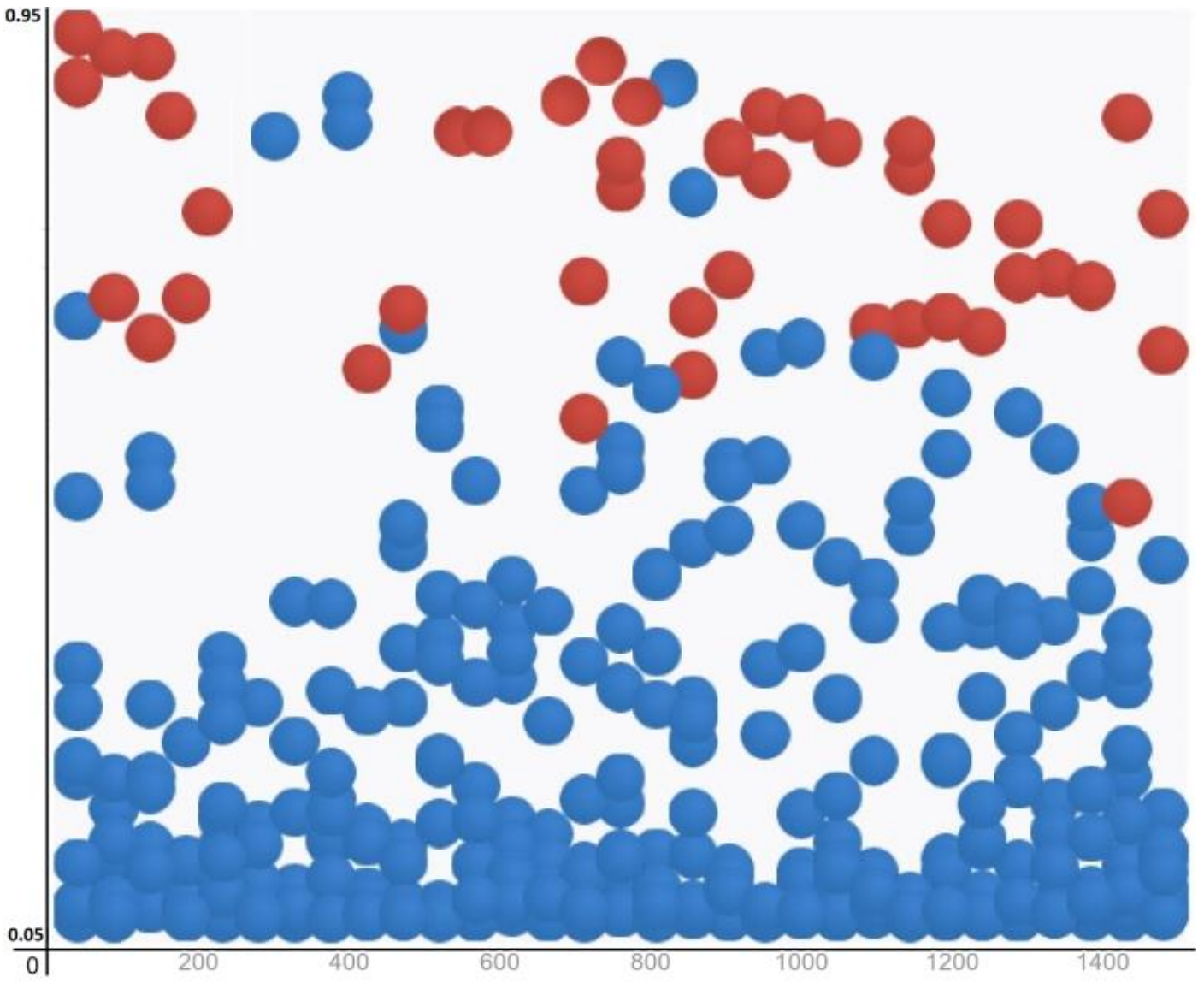

*Figure 13: False Negative (FN) steganography and cryptography masked content detection distribution in WAV (Red) and MP3 (blue) audio files.*

The FN rate in MP3 is much lower than the data that is identified accurately, but there is room for improvement. This analysis can be visualised in Figure 13. It is worth highlighting that the StegoHound toolkit includes a database creation script, and it could be expanded to loop through distinct types of data for further FN processing and improvement using file checksums and the use of an industry DB of blacklisted hashes which would significantly improve the FN reduction.

## 6. Conclusion and Discussions

The use of anti-forensics techniques has been democratised during the last decades and modern technologies were weaponized by cybercriminals to make LAE investigative tasks more complex and time-consuming. The results of this work reveal the impacts of combining different steganography detection techniques impact in enhancing the accuracy and efficiency of digital crime investigation, this research proposed a novel sequential and multi-approaches automation which is deemed as reliable to address the current gap in balancing the efficiency and cost (time and human labour) when it comes to steganography use detection at both law-enforcement and corporate which would avoid common cases of miscarriage of justice due to the complexity of the investigation. The proposed method's second contribution is covering more ground during digital forensics analysis and enhancing the results' reliability and thus overall case confidence. In the audio files context, detecting steganography in MP3 and WAV files can present several challenges due to the nature of steganographic techniques and the characteristics of audio files. Current cybercrime trends show a wider adoption of encryption and advanced techniques to mask and secure important messages and data within host files including audio files making it difficult to detect hidden information and causing Steganalysis Resistance, in practice steganography tools are being designed to resist detection by employing sophisticated methods that minimize statistical, audible, or perceptual changes in the audio data and therefore fooling most of industry digital forensics tools and frameworks. Finally, the proposed multi-approach automation validated the impact in terms of efficiency and accuracy notably on large audio files (MP3 and WAV) which the forensics analysis is time-consuming and also requires significant computational resources and memory but yet occasionally produces false positives (detecting steganography where none exists) or false negatives (failing to detect steganography that is present). Overall, the method tested in StegoHound demonstrated a good balance between detecting hidden data accurately and avoiding false alarms is a challenge. The adopted method for automating steganography detection and extraction is proven to be more efficient than existing tools from testing where existing tools were not capable of uncovering the data. The method implemented in the StegoHound is designed to be expandable due to the scripts' modular structure. In future works, additional steganographic techniques of audio and video file types could be added to produce a more versatile method focused mostly on the new identification techniques notably using AI and ML.